\documentclass[12pt,a4paper]{article}
\usepackage[english]{babel}
\usepackage{amsmath,amsthm}
\usepackage{amsfonts}
\usepackage{amssymb}
\usepackage{graphicx}

\title{RBA like problem with thermo--kinetics is non convex}
\author{M. Dinh and V. Fromion \\[2ex] {\footnotesize MaIAGE, INRA, Universit\'e Paris-Saclay, F-78350 Jouy-en-Josas, France}}

\date{}

\begin{document}

\maketitle

\begin{abstract}
The aim of this short note is to show that the class of problem involving kinetic or thermo--kinetic constraints in addition to the usual stoechiometric one is non convex.
\end{abstract}

\section{Introduction}
Computing efficiently and simultaneously the abundances of metabolites and proteins and the metabolic fluxes is a critical challenge in systems biology. A first problem formulation in 2009 resulted in a nonlinear non convex optimization problem \cite{MvBR:09}. However, it remains to determine if the non convexity is due to the problem itself or due to its formulation. In \cite{WHH:14,MRS:14}, the authors studied a constrained enzyme allocation problem with general enzyme kinetics in metabolic networks and mathematically proved that optimal solutions of the nonlinear optimization problem are elementary flux modes. Therefore, the computation of the optimal solutions is strongly related to the enumeration of elementary flux modes, which is computationally hard \cite{DyM:88} and untractable in practice for large metabolic networks \cite{KlS:02}. The result of \cite{WHH:14,MRS:14} suggest that the original problem composed of (a) a stoichiometric constraint on the metabolic fluxes; (b) a constraint on the allocation of proteins within the metabolic network; and (c) a kinetic constraint on the enzymatic capacity of the enzyme, is non convex.

In this note, we consider this problem in a simple case of the metabolic network. We show that there exist at least two local optima, proving that the class of problem is non convex.

\section{A non convex example}
\begin{figure}
  \centering
  \includegraphics[width=7cm]{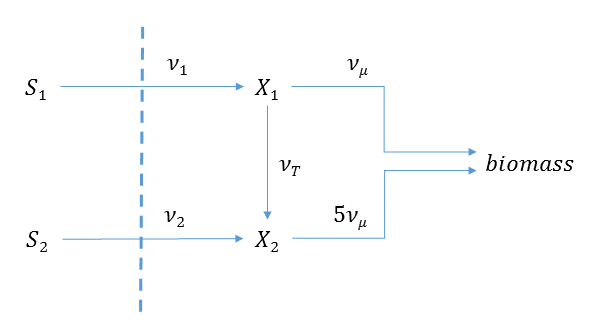}\\
  \caption{Metabolic network}\label{fig:metNet}
\end{figure}

We consider a metabolic network described by 3 reversible enzymatic reactions
$$
  \left\{
    \begin{array}{l}
      S_1 \stackrel{E_1}{\leftrightharpoons} X_1 \\
      X_1 \stackrel{E_T}{\leftrightharpoons} X_2 \\
      S_2 \stackrel{E_2}{\leftrightharpoons} X_2 \\
    \end{array}
  \right.
$$
as illustrated in Figure \ref{fig:metNet} and with the ``biomass reaction'' $X_1+5X_2 \overset{\nu_\mu}{\rightharpoonup} biomass$.
The metabolites $S_1$ and $S_2$ are external metabolites whereas $X_1$ and $X_2$ are internal ones. At steady state, the stoechiometric constraints are thus
$$
  \left\{
    \begin{array}{rcl}
      \nu_\mu & = & \nu_1 - \nu_T \\
      5 \nu_\mu & = & \nu_2 + \nu_T \\
    \end{array}
  \right.
  .
$$
The enzymatic fluxes are further constrained by thermodynamics and by the corresponding enzyme kinetics: we used the most plausible law proposed in \cite{LUK:10} with all the parameters of the law set to 1. This leads to further constraints
$$
  \left\{
    \begin{array}{l}
      \nu_1 = \frac{S_1-X_1}{1+S_1+X_1}E_1 \\[1ex]
      \nu_2 = \frac{S_2-X_2}{1+S_2+X_2}E_2 \\[1ex]
      \nu_T = \frac{X_1-X_2}{1+X_1+X_2}E_T \\
    \end{array}
  \right.
  .
$$

For this network, we want to maximize the biomass flux $\nu_\mu$, with a constraint on the enzymes concentration $E_1+0.1E_T+E_2\leq 1$. For given $S_1$ and $S_2$, this problem can be formulated as
$$
  \begin{array}{ll}
    \text{maximize} & \nu_\mu \\
    \multicolumn{2}{l}{\nu_1,\nu_T,\nu_2,\nu_\mu,E_1 \geq 0,E_T \geq 0,E_2 \geq 0,X_1 \geq 0,X_2 \geq 0} \\
    \text{under} & \\
      & \nu_\mu = \nu_1 - \nu_T \\
      & 5 \nu_\mu = \nu_2 + \nu_T \\
      & \nu_1 = \frac{S_1-X_1}{1+S_1+X_1}E_1 \\[1ex]
      & \nu_2 = \frac{S_2-X_2}{1+S_2+X_2}E_2 \\[1ex]
      & \nu_T = \frac{X_1-X_2}{1+X_1+X_2}E_T \\[1ex]
      & E_1+0.1E_T+E_3\leq 1 \\
  \end{array}
$$
and is an example of the problem considered in \cite{WHH:14,MRS:14}. The external concentrations are set to $S_1=0.17$ and $S_2=0.755$.

\bigskip

For this optimization problem, a brute force approach was implemented under Matlab 2012a using the {\tt fmincon} function and {\tt sqp} algorithm.
1000 random starts were performed with all decision variables between 0 and 1. It gives the solutions of Table \ref{tab:opt} and shows that there are at least 2 local optima corresponding to two elementary flux modes.
\begin{table}[hbtp]
\centering
\begin{tabular}{|l||c|c|}
  \hline
          &  A      & B       \\ \hline
  $\nu_1$ &  0      & 0.0540  \\
  $\nu_T$ & -0.0574 & 0       \\
  $\nu_2$ &  0.3442 & 0.2702  \\
  $\nu_\mu$ &  0.0574 & 0.0540  \\
  $E_1$   &  0      & 0.3719  \\
  $E_T$   &  1.0659 & 0       \\
  $E_2$   &  0.8934 & 0.6281  \\
  $X_1$   &  0      & 0       \\
  $X_2$     &  0.0569 & 0       \\
  \hline
\end{tabular}
\caption{Local optima} \label{tab:opt}
\end{table}

\bigskip

\begin{figure}
  \centering
  \includegraphics[width=15cm]{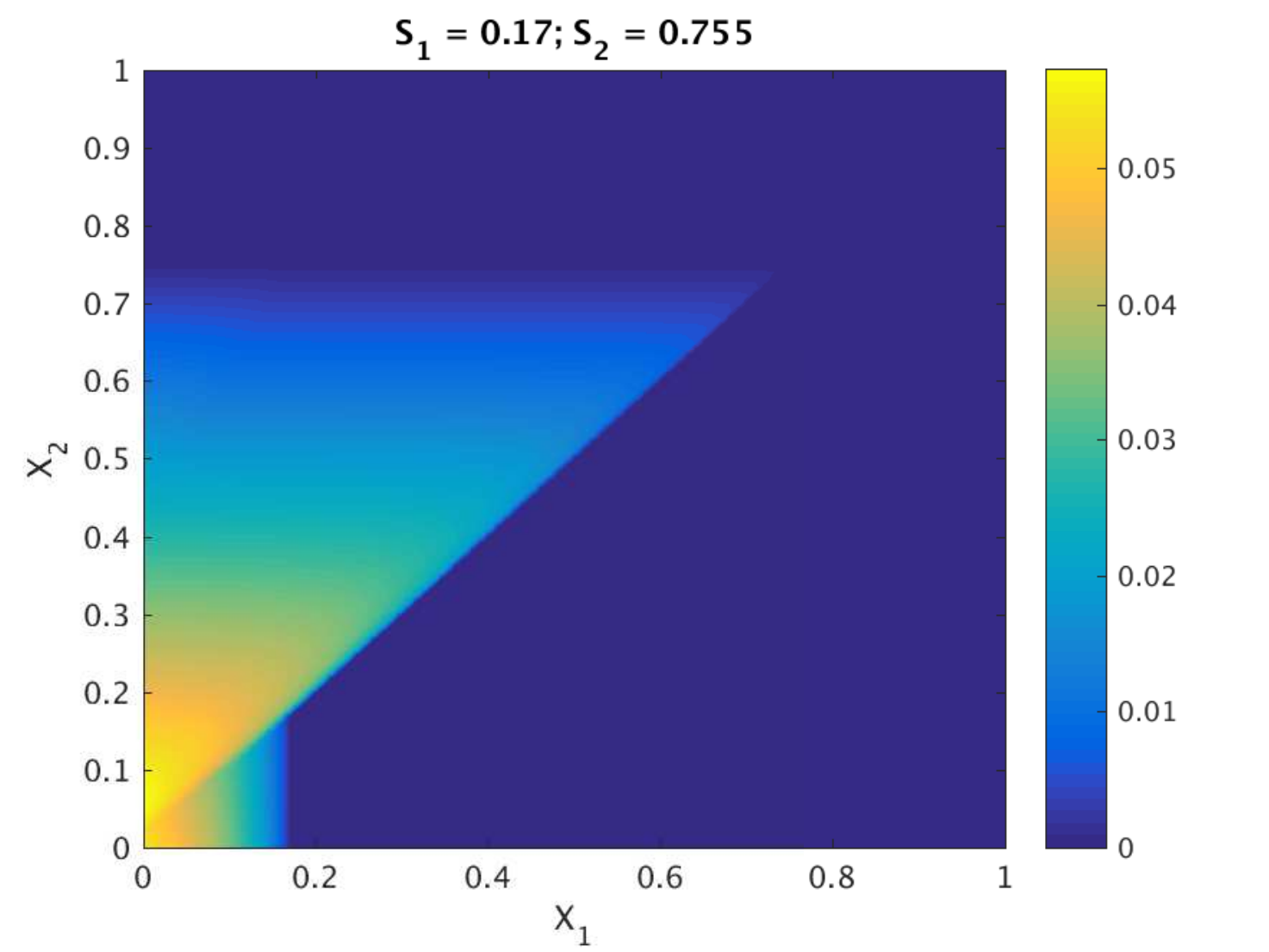}\\
  \caption{Representation of possible optima}\label{fig:nonConvexWhole}
\end{figure}

\begin{figure}
  \centering
  \includegraphics[width=15cm]{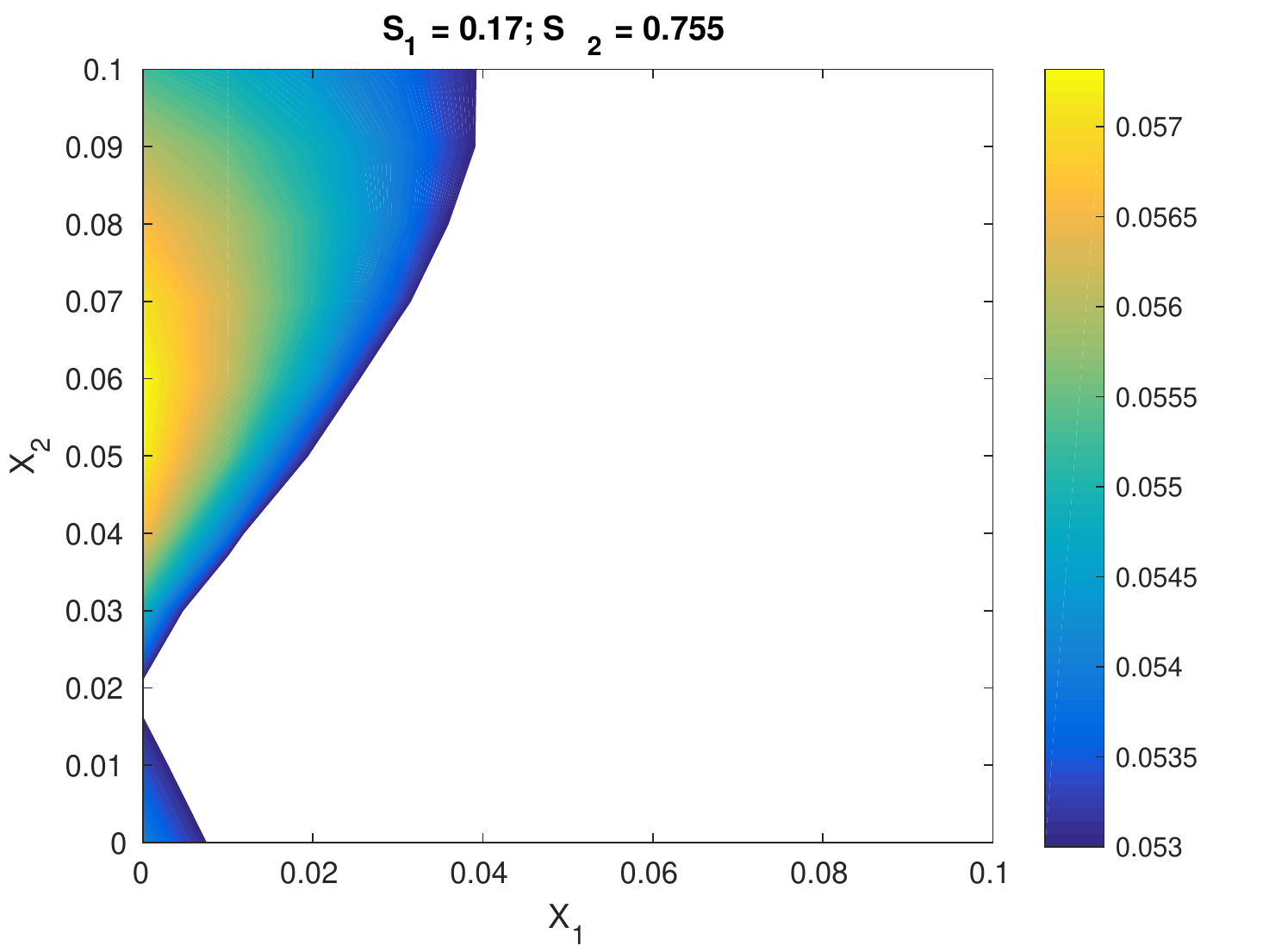}\\
  \caption{Representation of possible optima ('zoom')}\label{fig:nonConvex}
\end{figure}

Another way to obtain these optima is by noticing that, once $X_1$ and $X_2$ are set in the optimization problem, it becomes a Linear Programming problem so that the optimal is guaranteed to be obtained (we used {\tt cplex}). An illustration of these optima can be found in Figure \ref{fig:nonConvexWhole} and Figure \ref{fig:nonConvex} (please mind the different scales and colors between the figures). They represent the optimal flux $\nu_\mu$ as a function of the concentrations $X_1$ and $X_2$. The figures were obtained by gridding $X_1$ and $X_2$ (100 points for each between 0 and 1).

%

There exist at least two local optima for this example, proving that the considered class of problem is non convex.

\bibliography{dinhBio}

\begin{thebibliography}{1}

\bibitem{DyM:88}
M.E. Dyer and A.M. Frieze.
\newblock On the complexity of computing the volume of a polyhedron.
\newblock {\em SIAM Journal on Computing}, 17(5):967--974, 1988.

\bibitem{KlS:02}
S.~Klamt and J.~Stelling.
\newblock Combinatorial complexity of pathway analysis in metabolic networks.
\newblock {\em Molecular Biology Reports}, 29(1):233--236, 2002.

\bibitem{LUK:10}
W.~Liebermeister, J.~Uhlendorf, and E.~Klipp.
\newblock Modular rate laws for enzymatic reactions: thermodynamics,
  elasticities and implementation.
\newblock {\em Bioinformatics}, 26(12):1528--1534, 2010.

\bibitem{MvBR:09}
D.~Molenaar, R.~van Berlo, D.~de~Ridder, and B.~Teusink.
\newblock Shifts in growth strategies reflect tradeoffs in cellular economics.
\newblock {\em Molecular Systems Biology}, 5(1):n/a--n/a, 2009.

\bibitem{MRS:14}
S.~M{\"u}ller, G.~Regensburger, and R.~Steuer.
\newblock Enzyme allocation problems in kinetic metabolic networks: Optimal
  solutions are elementary flux modes.
\newblock {\em Journal of Theoretical Biology}, 347:182--190, 2014.

\bibitem{WHH:14}
M.T. Wortel, P.~Han, J.~Hulshof, B.~Teusink, and F.J. Bruggeman.
\newblock Metabolic states with maximal specific rate carry flux through an
  elementary flux mode.
\newblock {\em FEBS Journal}, 281(6):1547--1555, 2014.

\end{thebibliography}
\bibliographystyle{plain}

\end{document}